\documentclass[aps,prl,twocolumn,superscriptaddress,showpacs,amsmath,amssymb,longbibliography]{revtex4-1}
\usepackage[english]{babel}
\usepackage{amsmath}
\usepackage{bm}
\usepackage{graphicx,bbm} 
\usepackage{times}
\usepackage{epsfig} 
\usepackage[colorlinks,linkcolor=blue,citecolor=blue,urlcolor=blue]{hyperref}
\usepackage{color}
\usepackage{soul}
\definecolor{nred} {RGB}{224,0,0}
\definecolor{nblue} {RGB}{28,130,185}
\definecolor{dgreen} {RGB}{78,138,21}
\definecolor{norange}{RGB}{230,120,20}

\begin{document} 
\title{Delocalized charge carriers in strongly disordered $t$--$J$ model. }
\author{Janez Bon\v ca}
\affiliation{Jo\v zef Stefan Institute, SI-1000 Ljubljana, Slovenia}
\affiliation{Faculty of Mathematics and Physics, University of
Ljubljana, SI-1000 Ljubljana, Slovenia}
\author{Marcin Mierzejewski}
\affiliation{Institute of Physics, University of Silesia, 40-007 Katowice, Poland}
\begin{abstract} 
We study the influence of the electron--magnon interaction on the particle transport in strongly disordered systems.  
The analysis is based on results obtained for a single hole in the one--dimensional $t$--$J$ model. 
Unless there exists a mechanism that localizes spin excitations, the charge carrier remains delocalized even for a very strong disorder and shows subdiffusive motion 
up to the longest accessible times.  However,  upon inspection of the propagation times between neighboring sites as well as a careful finite--size scaling we conjecture that the anomalous subdiffusive transport
may be transient and should eventually evolve into a normal diffusive motion.
\end{abstract}
\maketitle

{\it Introduction.--} 
The many--body localization  (MBL)  represents a
promising concept of  macroscopic devices which do not thermalize \cite{pal10,serbyn13,lev14,schreiber15,serbyn15,khemani15,luitz16,ZZZ2_1,ZZZ2_2,ZZZ2_3,ZZZ2_4,ZZZ2_5,ZZZ2_6,ZZZ2_8,ZZZ2_9,ZZZ2_10,ZZZ2_11,JJJ3} and may store the quantum information \cite{ZZZ2_7,qi}. 
 Most of the inherent properties  of  MBL systems have been investigated using the generic one-dimensional (1D) disordered models of interacting spinless fermions\cite{ZZZ5_1,ZZZ5_2,ZZZ5_3,ZZZ5_4,ZZZ5_5,ZZZ5_6,ZZZ5_7,ZZZ5_8,ZZZ5_9,ZZZ5_10,ZZZ6_1}. Emerging characteristic features of MBL systems are: the existence of localized many-body states in the whole energy spectrum   that leads to vanishing of d.c. transport at any temperature \cite{berkelbach10,barisic10,agarwal15,gopal15,lev15,steinigeweg15,barisic16,kozarzewski16}, Poisson-like level statistics  \cite{oganesyan07}, and  the logarithmic growth of the entanglement entropy \cite{znidaric08,bardarson12,kjall14,serbyn15,luitz16,ZZZ4_1,ZZZ4_2}. 
Numerical calculations of dynamical conductivity \cite{agarwal15,steinigeweg15,barisic16} and  other dynamic properties  based on  the renormalization-group approach \cite{gopal15,vosk15,potter15,luitz16} indicate that  in the vicinity of the transition to MBL state the optical conductivity shows  a characteristic linear $\omega$-dependence.
In the presence of strong disorder but still below the MBL transition, several studies predict a subdiffusive transport \cite{agarwal15,gopal15,znidaric16,lastsub}.

The presence of  MBL has been  rigorously shown so far  only for the transverse--field Ising model \cite{imbrie16}, whereas the indisputable numerical evidence is available mostly for interacting spinless fermions 
or equivalent spin Hamiltonians.  
However, in real systems, the particles are coupled to other degrees of freedom and  this coupling may be important not only for solids but also for the cold--atom experiments. 
In particular,  the recent experiments  \cite{schreiber15,kondov15,bordia16}
address the problem of MBL in the spin--$1/2$ Hubbard model, where charge carriers are coupled to spin excitation.
On the other hand,  well established results \cite{basko06,mott1968} indicate that phonons destroy the Anderson localization,  hence they should destroy the MBL phase as well.
Nevertheless,  in contrast to phonons, the energy spectrum of many other excitations in the tight--binding models (e.g., the spin excitations)   is bounded from above. It remains  unclear whether strict MBL  survives in the presence of  the latter excitations. 
Solving this problem is important for answering the fundamental question whether MBL exists also in more realistic models including the Hubbard model \cite{rigol,peter}.  The preliminary numerical results suggest that charge carriers may indeed be localized despite the presence of delocalized spin excitations \cite{peter}.  Nonetheless, studies  for larger systems and longer times are needed in order to eliminate the  transient or finite-size (FS) effects.

We study dynamics of a single charge carrier  coupled to spin excitations which propagates in a disordered lattice.  Our studies are carried
out for the 1D $t$--$J$ model, which should be considered as a limiting case of the Hubbard model for large on-site repulsion.  The common understanding of MBL is that 
it originates from (a single--particle) Anderson insulator \cite{anderson58,mott68,kramer93,ZZZ1_1} which persists {\em despite} the presence of carrier-carrier interaction  
\cite{fleishman80,basko06}. The choice of a  single particle (hole) in the $t$--$J$ model eliminates the latter interaction, hence, it should act in favour of localization. Note however, 
that non--trivial many--body physics emerges from interaction between the spin excitations.  The dimension of the Hilbert space in the present studies is of the same order
as in the commonly studied model of spinless fermions, hence the numerical results are obtained for rather large systems and long times  far beyond the limitations  of the Hubbard model. 

We demonstrate that localization of charge carriers  is possible only for localized spin excitations, whereas  their dynamics  is subdiffusive even for very strong disorder (or diffusive for weak disorder) 
when spins are delocalized. The latter result resembles  the dynamics of  interacting spinless fermions for strong disorder but still  below the MBL transition
\cite{agarwal15,gopal15,znidaric16,lastsub}.  Here, we demonstrate that the subdiffusive behavior originates from extremely broad distribution of propagation--times for transitions between
the neighboring lattice sites.  However, this distribution strongly suggests that the subdiffusive behavior is a transient, yet long-lasting phenomenon.
The transition to the normal diffusive regime takes place at extremely long times and cannot be observed directly from numerical data. 

{\it Model and numerical methods.--}  We study a single hole (a charge carrier) in the 1D $t$--$J$ model on $L$ sites with periodic boundary conditions
\begin{eqnarray}
 H&=& \sum_{i,\sigma} [-t_{h} c_{i+1\sigma}^{\dagger} c_{i\sigma} + {\mathrm H.c}.  +\varepsilon_i  n_{i \sigma}] + \sum_{i}  J_i\vec{S}_{i+1} \vec{S}_{i},  \nonumber \\
\label{ham} 
\end{eqnarray} 
where $c^{\dagger}_{i\sigma} $ creates an electron with spin $\sigma$ at site $i$, $\vec{S}_{i}$  is the spin operator and $n_{i\sigma}=c^{\dagger}_{i\sigma}c^{}_{i\sigma}$.
The states with doubly occupied sites are excluded ($n_{i\uparrow} n_{i \downarrow}=0$) and, for simplicity,  the hopping integral is taken as the energy unit ($t_h=1$). 
The onsite  potentials,  $\varepsilon_i$, are random numbers that are uniformly distributed in the interval $[-W,W]$.  In the special case $J_i=0$,  the spins are frozen (at last in 1D case) and the hole dynamics should be the same as in the 1D  Anderson insulators.  

Considering the Hamiltonian (\ref{ham}) as a large--$U$ limit  of  the  Hubbard model, one finds that each $J_i$  depends on  $\varepsilon_i$ and   $\varepsilon_{i+1}$. However,  the disorder  in the Hubbard model always enlarges the exchange interaction \cite{distj},   $J_i \ge J(\varepsilon_j =\rm{const}) $, hence such disorder alone should not localize the spin excitations. In order to discuss also a more general case of localized spin excitations,  $J_i$  and  $\epsilon_i$ will be set independently of each other.   

The transport properties will be discussed mainly from the numerical results for the time propagation of pure states $|\psi_t \rangle$.
We take $|\psi_0 \rangle =| \sigma_1 ... \sigma_{j-1} 0_j \sigma_{j+1} ...\sigma_L \rangle$ as an initial state,  where  the position  of the hole, $j$,  as well as  the  spin configuration, $\sigma_i =\uparrow, \downarrow$, are chosen randomly. The latter choice means that the system is at infinite temperature.
The essential information on the charge dynamics will be obtained from  the hole density $\rho_i(t) =\langle \psi_t |1-n_{i\uparrow}  -n_{i\downarrow} |\psi_t \rangle$.
Then, one can define also the mean square deviation of the hole distribution \cite{robin2016}
\begin{equation}
\sigma^2(t)=\sum_i i^2 \rho_i(t) -\left[\sum_i i \rho_i(t) \right]^2.
\label{vari}
\end{equation}
 Throughout  the paper, the averaging over disorder will be marked by the subscript "d". We typically take $10^3$ realizations of the disorder.  
 
 {\em  The toy model ---}  In order to gain a deeper understanding of  the anomalous charge dynamics, numerical data for the $t$--$J$ model will be compared with results for a classical particle which
 randomly walks on the same 1D lattice.  As a toy model, we employ the continuous--time random walking in which particle waits for a time $\tau$ on each site $i$ before  jumping to the neighboring site $i-1$ or $i+1$. This model is well understood for various distributions of the waiting times $f(\tau)$ \cite{bouchaud}. In particular, if the average waiting time is finite, $\bar{\tau} =\int {\rm d} \tau\;  f (\tau) \tau < \infty $, the model shows  at long times normal diffusion, $\sigma^2(t) \propto t$. 
However, for a broad distribution of waiting times $f(\tau) \sim 1/\tau^{\mu+1}$ with $0 < \mu <1$, $\bar{\tau}$ becomes infinite and one obtains a subdiffusive transport  with
$\sigma^2(t) \propto t^{\mu}$, see   \cite{bouchaud,mon73}.

\begin{figure}[!htb]
\includegraphics[width=0.9\columnwidth]{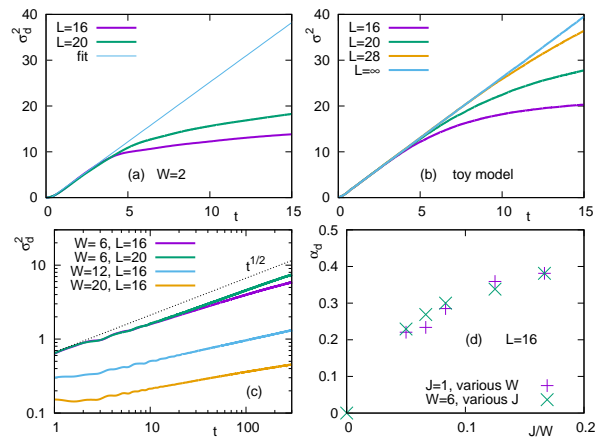}
\caption{a) and c) show the mean square deviation of the spatial distribution of holes vs. time obtained for $J_i=1$.
Note logarithmic scale in c). b) shows results for the toy model which reproduces the short--time linear regime in a).
d) shows $\alpha_d$ obtained from fitting $\sigma^2_d(t) \propto t^{\alpha_d}$  for  $L=16$ and various $J$ and  $W$. }
\label{fig1}
\end{figure}

{\em Results --} 
First we apply the Lanczos propagation method \cite{lantime} and study  weakly disordered system with homogeneous $J_i$, where one expects normal diffusion,  i.e., $\sigma^2_d(t) \propto t$. 
Such linear behavior is indeed visible in Fig. \ref{fig1}a  but only for short times. In order to explain the subsequent break--down of this linear trend  
 we have studied the toy model for exactly the same lattice  and $p (\tau) \propto \theta(\tau_0-\tau)$.  The average waiting time $\bar{\tau}=\tau_0/2$ has been tuned to fit the linear regime
in the $t$--$J$ model. Figures \ref{fig1}a and \ref{fig1}b show clear similarity between $\sigma^2(t)$ in both models. 
 However  in the toy model   any departure from the normal diffusion must originate  from the FS effects.  Due to these effects, the numerical results are of physical relevance only
 for small values of the mean square deviation $\sigma^2_d < \sigma^2_{\rm max} \sim 10 $.

In Figs. \ref{fig1}c and  \ref{fig1}d we present results for the central problem of this work, i.e., for the charge dynamics in strongly disordered $t$--$J$ model. 
The power--law dependence $\sigma^2_d(t) \propto t^{\alpha_d}$ is evident over  at least two decades of time and
the transport is clearly subdiffusive.
The exponent $\alpha_d \simeq 1/2$ for $W=6$ and further decreases for stronger disorder.  Within the studied time window $\sigma^2_d \ll \sigma^2_{\rm max}$. Therefore, we do not expect any essential influence of the FS effects. 
The latter hypothesis  is confirmed by numerical results  on larger system ($L=20$) where only slightly larger  value of the mean square deviation is obtained ($W=6$), shown also in Fig. \ref{fig1}c. 
Next, we have repeated the same calculations
for various (homogeneous) exchange interactions $J_i=J$  and various disorder strengths. For each case we have obtained  $\alpha_d$  and these exponents are shown  in  Fig. \ref{fig1}d 
as a function of $J/W$.  Nearly overlapping points on this plot suggest that $\alpha_d=\alpha_d(J/W)$. Then, the charge localization ($\alpha_d=0$) should occur  only for $W\rightarrow \infty$
or for $J \rightarrow 0$.  Localization is the former case is rather obvious, whereas the latter one is just the Anderson insulator.  Otherwise, the transport is subdiffusive or normal diffusive.

\begin{figure}[!htb]
\includegraphics[width=0.9\columnwidth]{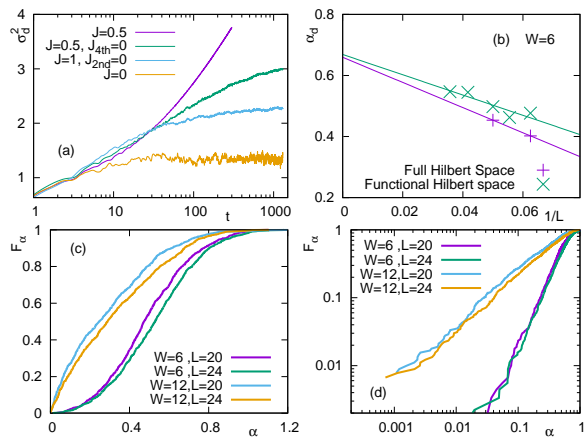}
\caption{a) shows $\sigma^2_d$ for $L=16$, $W=6$ and various spatial distributions of $J_i$. The latter quantity takes the value denoted by $J$ except for every 2nd or every 4th site $i$ where $J_i=J_{\rm 2nd}=0$ or  $J_i=J_{\rm 4th}=0$, respectively. b) shows $1/L$ scaling of the exponent $\alpha_d$ for homogeneous $J_i=1$.
c) and d) show cumulative distribution functions, Eq. (\ref{cuma}),  for  homogeneous $J_i=1$.  
Results in a)  are obtained from diagonalization of the Hamiltonian in the full Hilbert space while in c) and d) we used the functional space. } 
\label{fig2}
\end{figure}

The essential question is whether the model in Eq.~(\ref{ham}) may show  charge localization under some particular conditions.
In the following we demonstrate that such localization is indeed possible, provided 
that spin excitations are as well localized.  In order to localize the latter degrees of freedom we put $J_i=0$ for every 2nd or every 4th site $i$, otherwise we keep $J_i=1$.  Results are shown 
in figure \ref{fig2}a together with the data for the subdiffusive case ($J_i=J=0.5$) and the Anderson insulator ($J_i=J=0$).  Within  the time window that is accessible to our numerics,  we don't observe a complete saturation of  $\sigma^2_d(t)$ except for the Anderson insulator. However, since the increase is visibly slower than logarithmic we conclude that that hole is indeed localized.  An extremely slow charge dynamics within the localized regime is not very surprising.  It has previously been reported also for the MBL phase in a system of interacting spinless fermions \cite{our2016}.   
  
From now on, we study the details of  the subdiffusive transport in a system with homogeneous exchange interaction and, for simplicity, we set $J_i=1$.  Fig. \ref{fig2}b shows the exponents
$\alpha_d$ for various $L$. While the FS effects are not essential they are not negligible either. Therefore, it is important to employ a method which allows to study  even larger systems. 
In a case of single carrier instead of diagonalizing the Hamiltonian in the full Hilbert space we use the limited functional Hilbert space \cite{JJJ8}.
Such approach has successfully been applied to the studies on the real--time dynamics of $t$--$J$ and Holstein models \cite{JJJ5,JJJ6,JJJ7,JJJ9} and it is briefly explained also in the Supplemental material \cite{supp}.  
In this approach one accounts for all spin excitations in the closest vicinity of the holde but only for a selected more distant excitations. In contrast to the previous method, $L$ does not represent  the geometric size of the lattice but the maximal distance between the hole and the spin excitation. However, in both approaches one is interested in the limit $L \rightarrow\infty$ and the
corresponding  $1/L$ scaling of $\alpha_d$ is shown in Fig.   \ref{fig2}b.  Both methods obviously give the same extrapolated value of the exponent $\alpha_d$. However, diagonalization  in the Functional Hilbert space shows much weaker FS effects than the other approach. 

Next, we check whether possible isolated  cases with a localized hole have  been  overlooked  when discussing  results averaged over the disorder. 
To exclude the latter possibility,  we have fitted $\sigma^{2}(t) \propto t^{\alpha}$  independently for each realization of the disorder,  thus  generating  the distribution of the exponents $f(\alpha)$.  The calculations have been carried out for  times $t \le 10^3$. 
In Figures  \ref{fig2}c and \ref{fig2}d we show the cumulative distribution function, 
\begin{equation}
F_{\alpha}=\int_0^{\alpha} {\rm d} {\alpha'} f(\alpha'),
\label{cuma}
\end{equation}
which  vanishes for small $\alpha$ according to power law $\alpha^{\mu}$.  As shown in Fig. \ref{fig2}d, $\mu$ depends on the disorder  strength but  seems to be free from the FS effects. 
Therefore, we conclude that $F_{\alpha \rightarrow 0}=0$ also in the thermodynamic limit.  In  contrast, $F_{\alpha \rightarrow 0}=F_0>0$ would indicate localization. 
For  delocalized spin excitations in the $t$--$J$ model, the charge dynamics may be very slow but the hole is never localized at least not in the studied time window $(0,10^3)$. 

 It has recently been agued that  the SU(2)  symmetry precludes conventional MBL \cite{su21,su22,su23}. In order to explicitly show that the latter mechanism is not responsible
for delocalization of charge carriers in the present system, we have also considered the $t$--$J$ Hamiltonian with anisotropic spin--spin exchange interaction. Results in the Supplemental material \cite{supp}
show that breaking the SU(2) symmetry doesn't lead to  the charge localization even for very strong disorder $W=20$. 

Finally, we show that the hole dynamics in strongly disordered $t$--$J$ model may be qualitatively understood by studying the classical toy model. 
The properties of the latter  are determined by the distribution of  waiting times, hence one should first specify which quantity obtained for the $t$--$J$ model bears the closest resemblance to the classical  waiting time.  Since the toy model describes sequence of hoppings between the neighboring sites,  in the $t$--$J$ model  we define $\tau$  as 
the shortest time for which  the mean square deviation  (\ref{vari}) equals the lattice constant,  $\sigma^2(\tau)=1$.
Such  $\tau$ is well defined for each  realization of disorder, and one obtains the distribution of the waiting times $f(\tau)$ in the quantum model.
Since we are particularly interested in the  large--$\tau$ properties of $f(\tau)$,  we study the integrated distribution function
\begin{equation}
\int_{\tau}^{\infty} {\mathrm d} \tau'  f(\tau')=1-F_{\tau},
\label{cumt}
\end{equation}
where $F_{\tau}$is the cumulative distribution. For the algebraically decaying $f(\tau) \sim 1/\tau^{\mu+1}$ one  gets $1-F_{\tau} \propto 1/\tau^{\mu}$, where $\mu=1$  is the threshold value for the subdiffusive long--time behavior of the toy model.

Figure \ref{fig3}a shows the integrated distribution of the waiting times obtained in the $t$--$J$ model for the largest accessible systems and various disorder strengths. 
 This distribution closely follows predictions of the toy model.  In a system showing normal  diffusion ($W=2$) the distribution is very narrow, $1-F_{\tau}$ decays much faster than $1/\tau$ and the average  waiting time is quite short $\bar{\tau} \sim 1$.  In strongly disordered subdiffusive systems $1-F_{\tau}$ decays slower than $1/\tau$ and $\bar{\tau}$ should be very large, if not infinite.  
Therefore, our results strongly suggest that the subdiffusive transport originates from very broad  distribution of   the waiting times. 
A broad distribution of the propagation times   between the neighboring lattices sites  has its origin in   the disorder strength $W$ which
is by far the largest energy scale in the Hamiltonian. However, large disorder is frequently used in the studies of systems showing MBL.
Therefore, the present explanation of  the subdiffusive transport may apply also to other strongly disordered systems with many--body interactions \cite{agarwal15,gopal15,znidaric16} .

\begin{figure}[!htb]
\includegraphics[width=0.9\columnwidth]{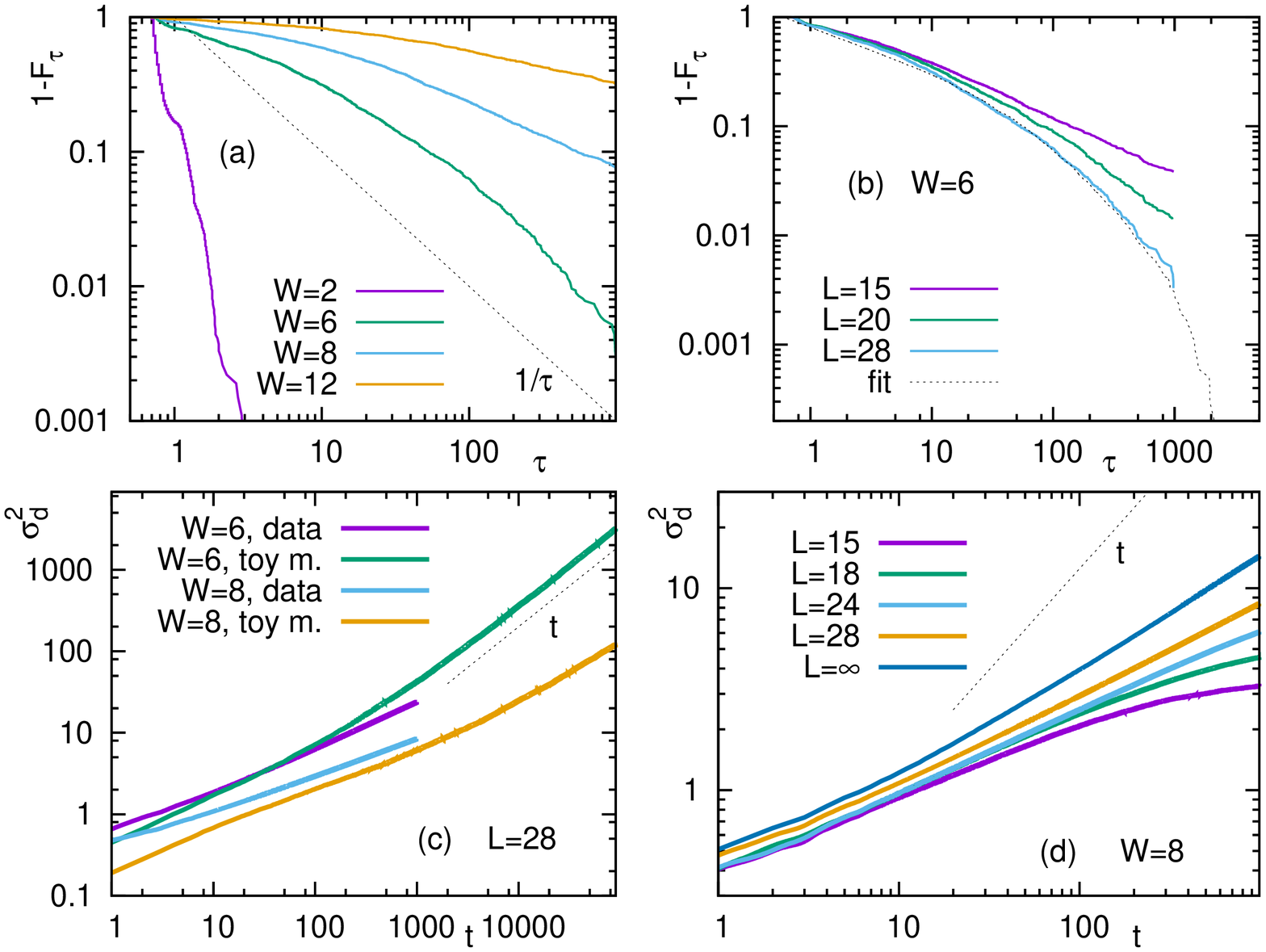}
\caption{ a)  shows integrated distribution of waiting times, Eq. (\ref{cumt}), for $L=28$ and various disorder strengths. 
The dashed  line marks the border between diffusive and subdiffusive regimes in the toy model.  b) shows the same but for $W=6$ and various $L$.  Here, the dashed
line shows the best fit. c) shows  $\alpha^2_d$ for the  $t$--$J$ model with $L=28$ (labelled as "data") and the toy model. 
The latter results have been obtained for the  distribution of waiting times shown in b) as as the best fit. d) shows  $\sigma^2_d$ for various $L$  together with the linearly extrapolated value $1/L \rightarrow 0$.       
Results have been obtained from diagonalization of the $t$--$J$ Hamiltonian in the limited functional Hilbert space.}
\label{fig3}
\end{figure}

The integrated distributions of waiting times shown in Figs. \ref{fig3}a and  \ref{fig3}b suggest that in the thermodynamics limit  and for sufficiently large $\tau$ the decay of 
$1-F_{\tau}$ may eventually become faster than $1/\tau$.  Then, the average waiting time will be huge but finite and the subdiffusive transport should be a long lasting yet transient  phenomenon. 
The time scale for the onset of the normal diffusion is far beyond the reach of any direct numerical studies of interacting quantum systems. 
However, such long--time regime can still easily be studied in the toy model.  In order to check this scenario, we have fitted  numerically obtained  results for the  waiting times of the $t$--$J$ model, as 
shown in  Fig.  \ref{fig3}b, and used this fit in the toy model. The resulting mean square deviation of the particle distribution is shown in  Fig.  \ref{fig3}c confirming the onset of normal
diffusion at $t \agt 10^4 $ for $W=6$ and  $t \gg 10^5 $ for $W=8$.

The conjecture with respect to   the transiency of the subdiffusive transport may also be supported by the analysis of     numerical results for the $t$--$J$ model without invoking the toy model.
As shown in Fig.  \ref{fig3}b , the width of the distributions of the waiting times decreases when $L$ increases. Therefore, one may expect that properly carried out FS  scaling 
may reveal  at least a clear tendency  for the transition to normal diffusion. 
In   Fig.  \ref{fig3}d we show results for $W=8$ and various $L$ together with  $\sigma^{2}_d(t)$ obtained from  linear in $1/L$ extrapolation to $1/L\rightarrow 0$. 
The slope of extrapolated curve gradually increases with $t$ already in the time--window which is accessible to our numerical procedure.

{\it Conclusions.--}  We have studied the dynamics of a single hole (charge carrier) in a strongly disordered $t$--$J$ model.  Our main result is that localization of the charge carriers should
be accompanied by localization of the spin--degrees of freedom, otherwise the charge dynamics is subdiffusive up to the longest times accessible to numerical calculations.  This holds true also for 
$t$--$J$--like Hamiltonians with broken  SU(2) symmetry.  However, based on the distribution of propagation times between the neighboring sites and after careful finite--size scaling of the mean square deviation we conjecture that the subdiffusive transport is transient and should eventually be replaced by a normal diffusion. According to the latter conjecture, the delocalized magnetic excitations 
in the thermodynamic limit become an infinite heat-bath which, similarly to electron-phonon coupling  \cite{basko06,mott1968}, restores non--zero albeit very small conductivity. 
While this conjecture requires further studies, the exceptionally broad distribution of propagation times  indicates that utmost care should be taken when formulating the claims on
the asymptotic dynamics based on numerical results obtained for times $\sim 10^3$ of the inverse hopping integrals.
 
 We expect that our qualitative claims should be valid also for other concentrations of charge carriers since each carrier is coupled to an infinite set of magnetic excitations, provided
that the latter excitations remain delocalized.  
However, an essential open problem is whether/which results reported here for the $t$--$J$ model remain valid also for the Hubbard model. 
Both models are equivalent provided that the Hubbard repulsion is stronger than all other energy scales including the disorder strength. Therefore, in strongly disordered systems the equivalence of
both models is restricted to very strong repulsions when the coupling between charge carriers and spin excitations is too weak to be studied with purely numerical methods.

{\it Acknowledgments.}
We acknowledge fruitful discussions with Fabian Heidrich-Meisner and Jerzy {\L}uczka. J.B. acknowledges the support by the program P1-0044 of the Slovenian Research Agency and
M.M. acknowledges support from the 2015/19/B/ST2/02856 project of the National Science Centre (Poland).

\bibliography{references}
\newpage

\large{\bf Supplemental Material}

\maketitle

\setcounter{figure}{0}
\setcounter{equation}{0}

\renewcommand{\thetable}{S\arabic{table}}
\renewcommand{\thefigure}{S\arabic{figure}}
\renewcommand{\theequation}{S\arabic{equation}}

\renewcommand{\thesection}{S\arabic{section}}

\label{pagesupp}

\label{sec:numerics}

In the Supplemental Material we study the charge dynamics in a disordered $t$--$J$--like model with broken SU(2) symmetry and  explain the construction of the functional Hilbert space.

{\it Hole dynamics in disordered, anisotropic $t$--$J$ model.--}  
Our main result concerns the diffusive or subdiffusive charge transport which persist in the $t$--$J$ model despite the presence of very strong disorder, $W\sim 20$,
unless the spin excitations are localized. It has recently been argued that the SU(2) symmetry precludes conventional many--body localization  \cite{su21,su22,su23}. 
Hence, it is important to check whether the SU(2) invariance  is responsible for the robust delocalized nature of  charge carriers also in the $t$--$J$ model. In order to answer this 
question, we study a modified version of the $t$--$J$ model
 \begin{eqnarray}
 H&=& \sum_{i,\sigma} [-t_{h} c_{i+1\sigma}^{\dagger} c_{i\sigma} + H.c.  +\varepsilon_i  n_{i \sigma}] \nonumber \\ 
 && + \sum_{i}  \left[ J (S^x_{i+1}  S^x_{i}+  S^y_{i+1}  S^y_{i})+  J^z S^z_{i+1}  S^z_{i} \right].  \nonumber \\ 
\label{hamsub} 
\end{eqnarray}  
When compared to the Hamiltonian (1) in the main text we have introduced anisotropic but site--independent exchange interaction such that 
the SU(2) invariance is broken for $J^z \ne J$. Otherwise, we use the same notation as in main text.

\begin{figure}[!htb]
\includegraphics[width=0.9\columnwidth]{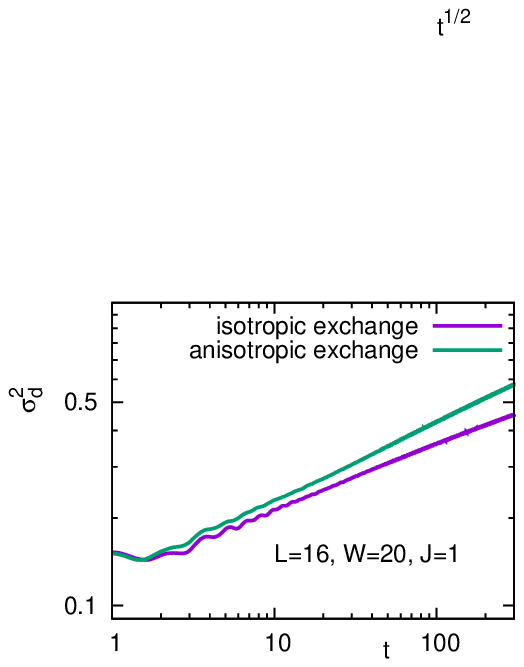}
\caption{Disorder--averaged mean square deviations [see Eq.(2) in the main text] vs. time for isotropic ($J^z=J=1$) and anisotropic ($J^z=2, J=1$) exchange interactions. }
\label{fig4sub}
\end{figure}

In Fig. \ref{fig4sub} we compare the disorder--averaged mean square deviations, $\sigma^2_d(t)$, obtained for isotropic ($J^z=J=1$) and anisotropic ($J^z=2, J=1$) systems from the Lanczos propagation method \cite{lantime}.   Despite a very strong  disorder, $W=20$,   breaking the SU(2)  symmetry doesn't lead to localization of charge carriers.
On the contrary, the subdiffusive transport $\sigma^2_d(t) \propto t^{\alpha_d}$ is clearly visible in both cases and the exponent $\alpha_d$ is even slightly larger in a system with broken SU(2) symmetry.
The latter result indicates that $\alpha_d$ is influenced by increasing $J^z$ rather than by breaking the SU(2) symmetry.

{\it Limited Functional Hilbert Space for the $t$--$J$ model.--}  
Generators of the Limited Functional Hilbert Space (LFHS) are derived from off-diagonal parts of the Hamiltonian (1) in the main text,
\begin{eqnarray}
O_1&=& \sum_{i,\sigma} c_{i+1\sigma}^{\dagger} c_{i\sigma} + H.c.  \nonumber \\ 
O_2&=& \sum_{i}  S^+_{i+1}  S^-_{i}+  S^-_{i+1}  S^+_{i}.
\end{eqnarray}
The generating algorithm starts from a hole at a given position, {\it e.g.} $i=0$ in a N\' eel state of spins, 
$\vert \psi^{(0)}\rangle = c_{0\sigma} \vert \mathrm{Neel}\rangle$.  We then apply the generator of basis $L$-times to generate the whole FHS:
\begin{equation} \label{lfhs}
\left\{|\psi^{(l)} \rangle \right\}=\left( O_1 + \tilde O_2\right) ^{l}|\psi_{(0)} \rangle,
\end{equation}
for $l=0,...,L$. The operator $\tilde O_2$ acts only on pairs of spins that due to hole motion deviate from the original N\' eel state.  $L$ represents the largest distance that the hole travels from its original position. In the case of LFHS we impose open boundary conditions. Sizes of LFHS span from $N_\mathrm{st}\sim 4000$ for $L=16$ up to  $5\times 10^5 $ for the largest $L=28$ used in our calculations. While even the largest size of LFHS seems rather small when performing exact-diagonalization procedures, this is not the case when performing time-propagation as well as sampling over more than 1000  samples. To achieve sufficient  accuracy of time propagation, we have used time-step-size  $\Delta t=0.02$ and performed up to $5\times 10^5 $ time steps. The advantage of LFHS over the exact diagonalization approach  is to significantly reduce the Hilbert space by generating  spin excitations in the proximity of the hole, which  in turn allows for  investigations of larger system sizes. After completing generation of LFHS we time evolve Hamiltonian in Eq.~(1)  while taking the advantage of the  standard Lanczos-based diagonalization technique. The finite-size scalings of various quantities with increasing $L$ are presented in Figs. 2b, 3b, and 3d in the main text. The method has been successful in computing static and dynamic properties  \cite{JJJ8} as well as non-equilibrium dynamics \cite{JJJ7,JJJ6,JJJ5} of correlated electron systems.

\end{document}